\documentclass[preprint,11pt,authoryear]{elsarticle}

\usepackage{caption}
\usepackage{subcaption}
\usepackage{graphicx}
\usepackage{xcolor}
\usepackage{balance}
\usepackage{url}
\usepackage{amsmath}
\usepackage{mathtools}
\usepackage{array,relsize}
\usepackage{ragged2e}
\usepackage{algorithm}
\usepackage[noend]{algpseudocode}
\usepackage{float}
\usepackage{multirow}
\usepackage{todonotes}
\usepackage[export]{adjustbox}
\usepackage{soul}
\usepackage[T1]{fontenc}
\usepackage[utf8]{inputenc}
\usepackage{babel}
\usepackage{footnote}
\usepackage{lineno}
\usepackage{csquotes}
\usepackage{multicol}
\usepackage{natbib}
\usepackage{eso-pic}

\makesavenoteenv{tabular}
\makesavenoteenv{table}

\algdef{SE}[VARIABLES]{Variables}{EndVariables}
   {\algorithmicvariables}
   {\algorithmicend\ \algorithmicvariables}
\algnewcommand{\algorithmicvariables}{\textbf{global variables}}

\algnewcommand\algorithmicglobals{\textbf{Globals:}}
\algnewcommand\GLOBALS{\item[\algorithmicglobals]}
\algrenewcommand\algorithmicrequire{\textbf{Input: }}
\algrenewcommand\algorithmicensure{\textbf{Output: }}

\journal{Journal of Cleaner Production}

\begin{document}

\begin{frontmatter}


\title{SEER: Sustainable E-commerce with Environmental-impact Rating}

\author[1]{Md Saiful Islam}
\ead{mislam6@ur.rochester.edu}
\address[1]{Department of Computer Science, University of Rochester, United States}
\author[1]{Adiba Mahbub Proma}
\ead{aproma@ur.rochester.edu}
\author[1]{Caleb Wohn}
\author[2]{Karen Berger}
\address[2]{Department of Earth and Environmental Sciences, University of Rochester, United States}
\author[1]{Serena Uong}
\author[1]{Varun Kumar}
\author[3]{Katrina Smith Korfmacher}
\address[3]{Department of Environmental Medicine, University of Rochester Medical Center, United States}
\author[1]{Ehsan Hoque\corref{ca}}
\cortext[ca]{Corresponding Author}
\ead{mehoque@cs.rochester.edu}
      
\date{}

\begin{abstract}
With online shopping gaining massive popularity over the past few years, e-commerce platforms can play a significant role in tackling climate change and other environmental problems. In this study, we report that the ``attitude-behavior'' gap identified by prior sustainable consumption literature also exists in an online setting. We propose SEER, a concept design for online shopping websites to help consumers make more sustainable choices. We introduce explainable environmental impact ratings to increase knowledge, trust, and convenience for consumers willing to purchase eco-friendly products. In our quasi-randomized case-control experiment with 98 subjects across  the United States, we found that the case group using SEER demonstrates significantly more eco-friendly consumption behavior than the control group using a traditional e-commerce setting. While there are challenges in generating reliable explanations and environmental ratings for products, if implemented, in the United States alone, SEER has the potential to reduce approximately 2.88 million tonnes of carbon emission every year.
\end{abstract}


\end{frontmatter}

\section{Introduction} \label{sec:introduction}
In recent years, the effects of climate change and environmental pollution have become devastating, affecting communities worldwide. While awareness is increasing among the general population, many people lack the knowledge or motivation to make sustainable choices. This is specifically evident in consumer buying behavior -- despite individuals being concerned about the environment and willing to opt for greener consumption, the intentions are often not translated into appropriate actions \citep{young2010sustainable}. This phenomenon, termed as the ``attitude-behavior gap'', is commonly identified in consumption behavior literature \citep{tanner2003promoting, vermeir2006sustainable, wheale2007ethical}. High prices, difficulty in identifying green products, not having enough time for research, lack of environmental information in the product description, and lack of trust in the ``eco-friendly'' labels provided by the manufacturers have been identified as the contributing factors to the attitude behavior gap\citep{gleim2013against, joshi2015factors}. Individual and social factors like habit, behavioral control, social norms, and so on also contribute to the phenomenon, but their impacts are not well established \citep{eze2013green, tsakiridou2008attitudes, wang2014factors}. While the factors responsible for the attitude-behavior gap are well studied, existing literature falls short when providing viable solutions or strategies to mitigate this gap. 

\begin{figure}[ht]
\centering
\includegraphics[width=0.95\linewidth]{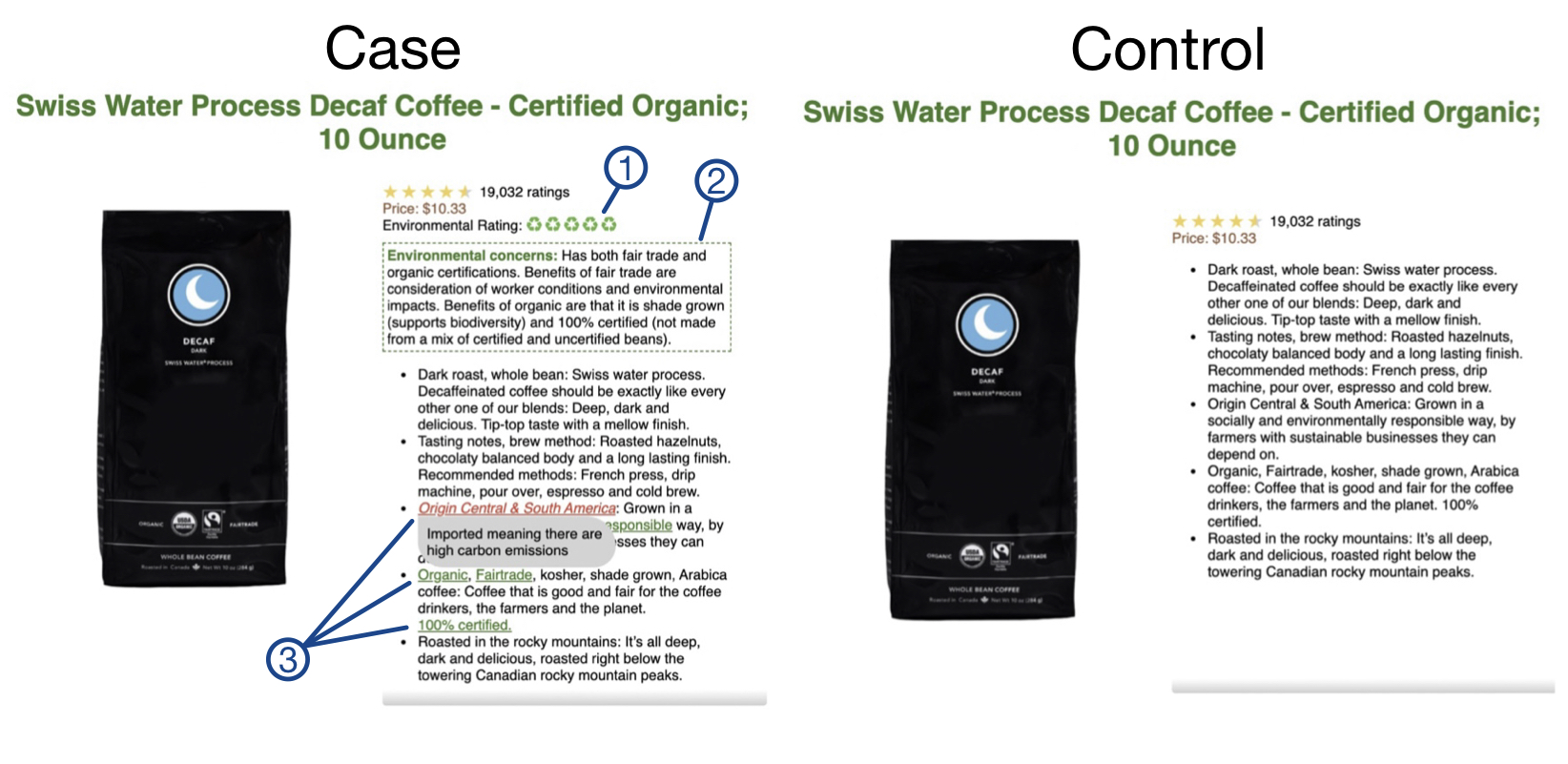}
\caption{(Left) shows our proposed design for displaying a product used by the case group. The \emph{environmental rating} \textbf{(1)} indicates how eco-friendly a product is and the \emph{environmental concerns} \textbf{(2)} summarizes the possible environmental impacts of the product. Keywords related to environmental impact analysis are also highlighted with green (eco-friendly) or red (harmful to the environment) color \textbf{(3)}. If a user hovers the mouse/pointer over the highlighted keywords, additional environmental information about the keywords is presented. (Right) shows the traditional product display design presented to the control group. In both cases, brand names and logos are removed from all products to avoid bias towards popular brands.}
\label{fig:main_example}
\end{figure}

Prior research suggests that green consumers prefer being presented with environmental impact information in a simple, compact, and user-friendly manner \citep{mondelaers2009importance}. At the same time, they often find it hard to trust ``eco-friendly'' labels provided by companies \citep{chen2012enhance}. The increasing popularity of online shopping brings a unique opportunity - the architecture of the website makes it easy to present information through the incorporation of high-resolution images, videos, and hypertexts that can point users to further resources. E-commerce websites can facilitate additional information about a product and present it in a way that does not overwhelm consumers but is still trustworthy and well-explained. In addition, whether an additional piece of information should be presented to a consumer can be customized based on consumer preference and behavior analysis. We find this as an excellent opportunity to convey the required information about a product's environmental impact, and aid users in making more eco-friendly decisions. For example, when users search for a specific product, they can sort the results based on the eco-friendliness rating. However, the question remains -- would consumers trust these ratings? To build trust, an explanation for the rating must be provided. There can be a summary of the product's environmental impacts that explains the rating of the product. In addition, from the description of the product, words or phrases (keywords) that are related to environmental impact can be highlighted and explained. This would also increase consumers' environmental knowledge, as they learn about the environmental impacts of various product ingredients, thereby raising public awareness. In this paper, we analyze the ``attitude-behavior gap'' for an online setting and propose \textbf{SEER} (\textbf{S}ustainable \textbf{E}-commerce with \textbf{E}nvironmental-impact \textbf{R}ating), a concept design (Figure \ref{fig:main_example}) that would facilitate making eco-friendly choices by incorporating the features discussed above.


We conduct a quasi-randomized case-control study of 98 participants, where participants are asked to select products for a local school with a limited budget. Before the start of the study, the participants are shown a climate awareness video and how their choices can make a difference. To elicit real-life behavior, everyone gets financial rewards for saving budget while selecting the products. The participants then complete a post-study survey related to whether their attitude towards the environment has changed (both groups) and the design interface of the website (case group only). From our observational study, we identify that the ``attitude-behavior gap'' also exists in an online setting -- participants who are more concerned about the environment do not choose a higher number of eco-friendly products. However, providing more information about the environment through our interface makes a difference. We find that the purchasing behavior of \emph{case} subjects using SEER differs significantly from the \emph{control} subjects using a traditional e-commerce website setting ($p < 0.01$). The case group demonstrates more eco-friendly behavior than the control group ($p < 0.005$) and reports that the components introduced in SEER made finding eco-friendly products more convenient, simultaneously increasing their trust in the labels because of the provided explanation. At the same time, SEER has been rated as nearly excellent in terms of usability, as established by a score of $79.18$ in System Usability Scale \citep{brooke1996sus}.

In 2020, the worldwide e-retail sale was more than 4.2 trillion USD with over two billion e-commerce customers \citep{coppola2021commerce}.  According to the quarterly sales report of the U.S. Census Bureau, e-commerce sales in the second quarter of 2021 accounted for 12.5 percent of total sales \citep{estquarterly2021}. This shows that,  despite the challenges of generating reliable environmental ratings and explanations, our proposed idea can have a huge impact in tackling climate change by potentially reducing harmful environmental effects from every online purchase. Adaptation of SEER will raise general awareness about climate change and individual responsibility, thus promoting sustainable products and encouraging the industry to invest more in sustainable production. Therefore, SEER can be the seer of a greener future. 



\section{Study Design} \label{sec:method}

\subsection{Interface design}


In this study, we develop a prototype website named \textbf{SEER} (\textbf{S}ustainable \textbf{E}-commerce with \textbf{E}nvironmental-impact \textbf{R}ating) that targets addressing three major factors responsible for the ``attitude-behavior'' gap observed in sustainable consumption literature: (i) inconvenience  (ii) lack of knowledge, and (iii) lack of trust. The key component of SEER is an \emph{environmental rating} (1-5 scale) that rates a product based on its environmental impact, with a higher rating indicating higher eco-friendliness of a product. However, for facilitating green consumption behavior, both the presence of a singular rating and its credibility are important \citep{riskos2021ecolabels, young2010sustainable}. Therefore, our second proposed component is an \textbf{environmental concerns} statement which briefly conveys an explanation of the rating provided by discussing the product's potential impact on the environment. These explanations can help to increase trust \citep{pu2006trust}. At the same time, we try to educate people by highlighting words or phrases (keywords) that are related to the environment, simultaneously explaining what these keywords mean. This is our proposed third component, named as \textbf{environmental keyword highlights}.

\subsection{Participant recruitment and pre-study survey}
We use Amazon Mechanical Turk \citep{paolacci2010running} to recruit 98 participants living in the United States. We run a case-control study by splitting the participants into two groups: case group selecting products using SEER ($N = 49$), and control group using a prototype of a traditional website ($N = 49$). Figure \ref{fig:main_example} shows both of the websites -- although the look and feel of both prototypes are the same, SEER additionally incorporates the three components proposed above. 

At first, the participants are asked to fill out some demographic information such as age range, ethnicity, gender, household income, the highest degree of education, and zip code. Then they are asked a standard set of questions used in prior literature to infer their environmental sentiment (i.e., how concerned they are about the environment how actively they are trying to prevent environmental harm) \citep{lin2012influence}. Depending on their answers, they are assigned an ``environmental sentiment score'' (0-40). The participants scoring at least 20 are considered sensitive or concerned about the environment, and a quasi-randomized process is used to assign a participant to either case or control group in such a way that both groups have a similar distribution in terms of the environmental sentiment score. In addition, we ask 5 standard questions to infer the knowledge of the participants about environmental issues and climate change. All the questions are on a 5-point Likert scale (strongly disagree: 0, disagree: 1, neutral: 2, agree: 3, strongly agree: 4). The quasi-randomized assignment is successful, as we observe similar distribution in both case and control groups -- not only for environmental sentiment score but also for gender, age, race, education, environmental knowledge, and income level. The detailed demographic distribution of the participants is presented in Table \ref{tab:demography}.

\begin{table}[]
\begin{tabular}{|l|l|l|l|l|}
\hline
\textbf{Variables}                                                                     & \textbf{Options}                                                    & \textbf{N (number)} & \textbf{Case (\%)} & \textbf{Control (\%)} \\ \hline
\multirow{2}{*}{Gender}                                                                & Female                                                                & 28                  & 46.4\%             & 53.6\%                \\ \cline{2-5} 
                                                                                       & Male                                                              & 70                  & 51.4\%             & 48.6\%                \\ \hline
\multirow{6}{*}{Age}                                                                   & 18-24                                                               & 4                   & 100.0\%            & 0.0\%                 \\ \cline{2-5} 
                                                                                       & 25-34                                                               & 47                  & 46.8\%             & 53.2\%                \\ \cline{2-5} 
                                                                                       & 35-44                                                               & 25                  & 40.0\%             & 60.0\%                \\ \cline{2-5} 
                                                                                       & 45-54                                                               & 12                  & 58.3\%             & 41.7\%                \\ \cline{2-5} 
                                                                                       & 55-64                                                               & 7                   & 71.4\%             & 28.6\%                \\ \cline{2-5} 
                                                                                       & 65 and above                                                        & 4                   & 25.0\%             & 75.0\%                \\ \hline
\multirow{4}{*}{Race}                                                                  & Asian                                                               & 7                   & 57.1\%             & 42.9\%                \\ \cline{2-5} 
                                                                                       & \begin{tabular}[c]{@{}l@{}}Black or\\ African American\end{tabular} & 12                  & 41.7\%             & 58.3\%                \\ \cline{2-5} 
                                                                                       & White                                                               & 78                  & 50.0\%             & 50.0\%                \\ \cline{2-5} 
                                                                                       & Other                                                               & 1                   & 100.0\%            & 0.0\%                 \\ \hline
\multirow{4}{*}{Education}                                                             & Below High School                                                   & 1                   & 0.0\%              & 100.0\%               \\ \cline{2-5} 
                                                                                       & Associate                                                           & 7                   & 57.1\%             & 42.9\%                \\ \cline{2-5} 
                                                                                       & Bachelor                                                            & 54                  & 44.4\%             & 55.6\%                \\ \cline{2-5} 
                                                                                       & Graduate                                                            & 15                  & 66.7\%             & 33.3\%                \\ \hline
\multirow{6}{*}{\begin{tabular}[c]{@{}l@{}}Yearly Family \\ Income (USD)\end{tabular}} & Below 20,000                                                        & 11                  & 45.5\%             & 54.5\%                \\ \cline{2-5} 
                                                                                       & 20,000-35,000                                                       & 13                  & 53.8\%             & 46.2\%                \\ \cline{2-5} 
                                                                                       & 35,000-50,000                                                       & 16                  & 56.3\%             & 43.7\%                \\ \cline{2-5} 
                                                                                       & 50,000-75,000                                                       & 34                  & 52.9\%             & 47.1\%                \\ \cline{2-5} 
                                                                                       & 75,000-100,000                                                      & 30                  & 36.7\%             & 63.3\%                \\ \cline{2-5} 
                                                                                       & Above 100,000                                                       & 5                   & 40.0\%             & 60.0\%                \\ \hline
\multirow{1}{*}{\textbf{Total}} & \textbf{-} & \textbf{98} & \textbf{50.0\%} & \textbf{50.0\%} \\ \hline

\end{tabular}
\caption{Demographic details of the study participants}
\label{tab:demography}
\end{table}


\subsection{Experiment setup}

\begin{figure}[ht]
\centering
\includegraphics[width=\columnwidth]{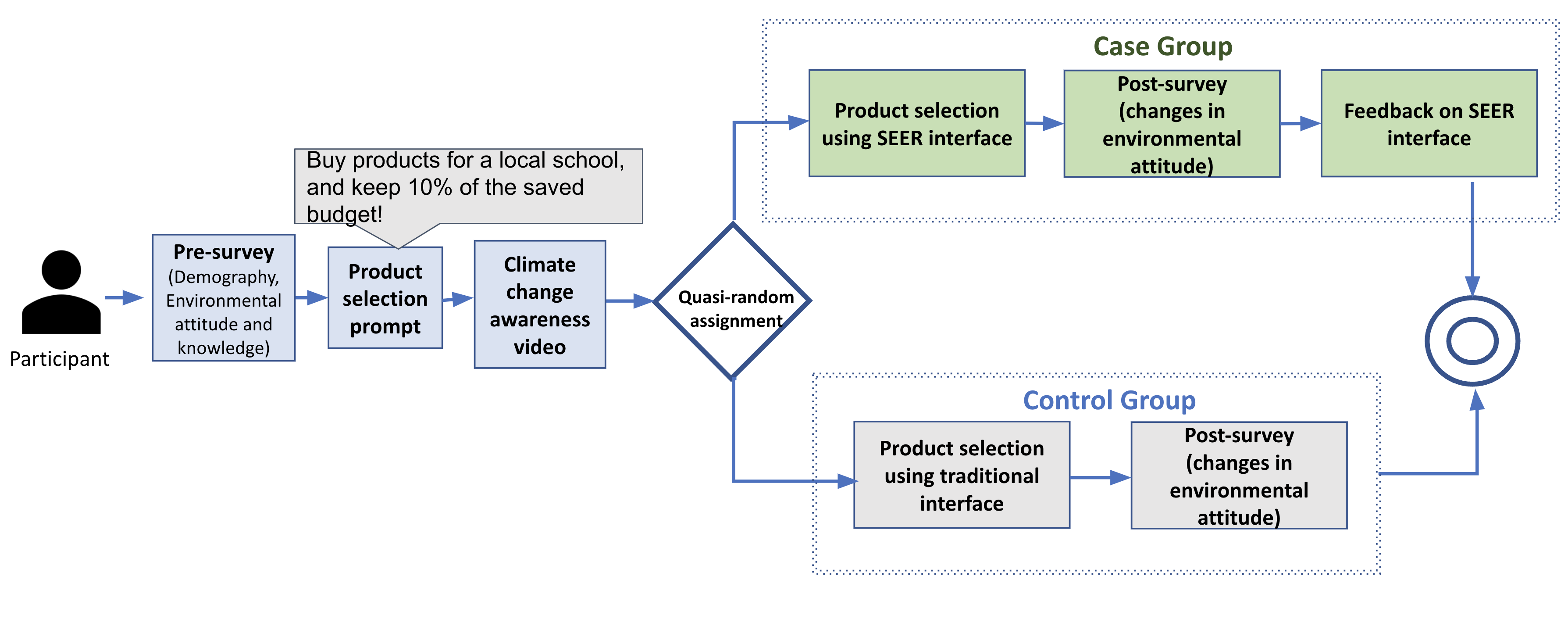}
\caption{A simple overview of our study design.}
\label{fig:overview}
\end{figure}

Figure \ref{fig:overview} summarizes a participant's activity in our study. Both case and control groups are given the prompt that a local school is just opening, and the school needs help to purchase some products due to a shortage of school staff (note that although the school is imaginary, participants are not aware of this until the end of the experiment). The prompt includes that the school has a limited budget, and if participants can save some money from the budget, they would receive 10\% of the saved money as a token of thanks. All participants are also shown a motivational video (3-minute length) explaining the adverse effects of environmental pollution and climate change, and how their individual actions can make a difference. The objective of showing this video is to prime the participants from both case and control groups into being more concerned about the environment and indirectly elicit eco-friendly behavior. In addition, showing the video helped us avoid adding any possible confounding factor between case and control groups. Since the case group would be provided with information on environmental impact, they are more likely to infer that the study is about making eco-friendly choices. Alternatively, the control group would have no such information, thereby going into the study with a completely different thought process. Priming both case and control groups with a motivational climate video removes this bias. Therefore, participants would have two possibly conflicting objectives in mind: (i) save money and get monetary benefits instantly, and/or (ii) invest in eco-friendly products which has no short-term benefits but will be helpful for the environment in the long run.

All participants are then shown a pair of products side-by-side from 12 pre-selected product types. Between each pair, one is more eco-friendly than the other. Figure \ref{fig:main_example} shows how each product was presented to the participants. The products are arranged in the same order, and in the same arrangement for all participants. Whether the eco-friendly product from a certain pair will be displayed on the left side or right side was pre-determined randomly (but kept the same for all participants). This avoids possible confounding factors (e.g., users being biased towards the product on the left side, being more attentive at the beginning, etc.). When selecting a product from a pair, the participants from the case group are shown three additional components (see Figure \ref{fig:main_example}) proposed in SEER -- (i) environmental rating, (ii) environmental concerns, and (iii) environmental keyword highlights, along with traditionally available features: product name, an image of the product, price, user rating, number of users who rated this product, and product description. The control group has no access to SEER components but saw the traditional features.

After the participants finish product selection, they complete a post-study survey. Both case and control groups answer the same questions related to climate sentiment as before. This time, however, the questions are asked in the future tense to determine the effectiveness of our intervention strategy in raising awareness of climate change and increasing concern for the environment among the participants. The participants from the case group answer additional questions regarding the effectiveness of the SEER components, system usability \citep{brooke1996sus}, and their experience (i.e., likes, dislikes, and suggestions) of the study.




\subsection{Products and their environmental impact assessment}
Based on the theme, we select 12 types of products displayed in Table \ref{tab:product_types} to be included in the study. We choose two products for each type, one being more eco-friendly than the other. For selecting the products and generating their relevant environmental information, we consult a domain expert (an Associate Professor of Environmental Science who is actively engaged in several sustainability programs). 

We remove the name and logo of the brand of the products since consumers can be biased towards certain brands \citep{chovanova2015impact, hillenbrand2013better, mellens1996review}. We also consider the price of the products as an additional context, and selected products to evenly represent each of the four different conditions (3 product pairs from each condition): 

\begin{itemize}
\item[(i)] Eco-friendly product is cheaper 
\item[(ii)] Both products are equally priced 
\item[(iii)] Eco-friendly product is slightly more expensive (additional cost is below 5 USD)
\item[(iv)] Eco-friendly product is significantly more expensive (additional cost is at least 5 USD)
\end{itemize}

These conditions would allow us to probe into the question of how strong the influence of price is for consumers when buying eco-friendly products.

\begin{table}[]
\centering
\begin{tabular}{|l|l|}
\hline
\textbf{\begin{tabular}[c]{@{}l@{}}Product Type \\ Code\end{tabular}} & \textbf{Product Type} \\ \hline
P1                         & Decaf coffee          \\
P2                         & All-purpose cleaner   \\
P3                         & Toilet paper          \\
P4                         & Clubhouse playset     \\
P5                         & 4-in-a-row board game \\
P6                         & Copy paper            \\
P7                         & Kitchen trash bags    \\
P8                         & Ballpoint pen         \\
P9                         & Chair                 \\
P10                        & Table                 \\
P11                        & Glue sticks           \\
P12                        & File folder           \\ \hline
\end{tabular}
\caption{12 types of products selected for the study}
\label{tab:product_types}
\end{table}

The domain expert rated the eco-friendliness of the products on a scale of 1 to 5 (higher value means more eco-friendly), and provided justifications for the rating. This justification is used as the ``environmental concerns'' in SEER. Furthermore, three undergraduate students highlighted keywords in the product description in either red or green, where red indicated a keyword that is not eco-friendly (e.g., pollution), green being the opposite (e.g., recycled). Additional justification as to why these keywords are marked in red or green and are also available to the case group upon hovering over the highlighted keywords.

\section{Results} \label{sec:results}


Participants from both groups demonstrate similar distribution in terms of other factors like race, gender, education, income, and environmental knowledge (Table \ref{tab:demography}). We study statistical correlations between independent variables (gender, age range, level of education, yearly income range, environmental sentiment score, and knowledge) and the target variable (eco-friendliness). Gender, age range, level of education, yearly income range are categorical values, while environmental sentiment score and knowledge are numerical. For variables with categorical values, we use the ANOVA test and report the $p$-values wherever appropriate. For variables with numerical values, we use the Pearson correlation co-efficient and report $r$. Also, to test the major hypotheses, we use the $t$-test and report $p$-values. System Usability Scale (SUS) score is measured exactly as in literature \citep{brooke1996sus}.

\subsection{People have good intention}
\begin{figure}%
    \centering
    \subfloat[\centering]{\label{fig:climate_senti}{\includegraphics[width=0.45\columnwidth]{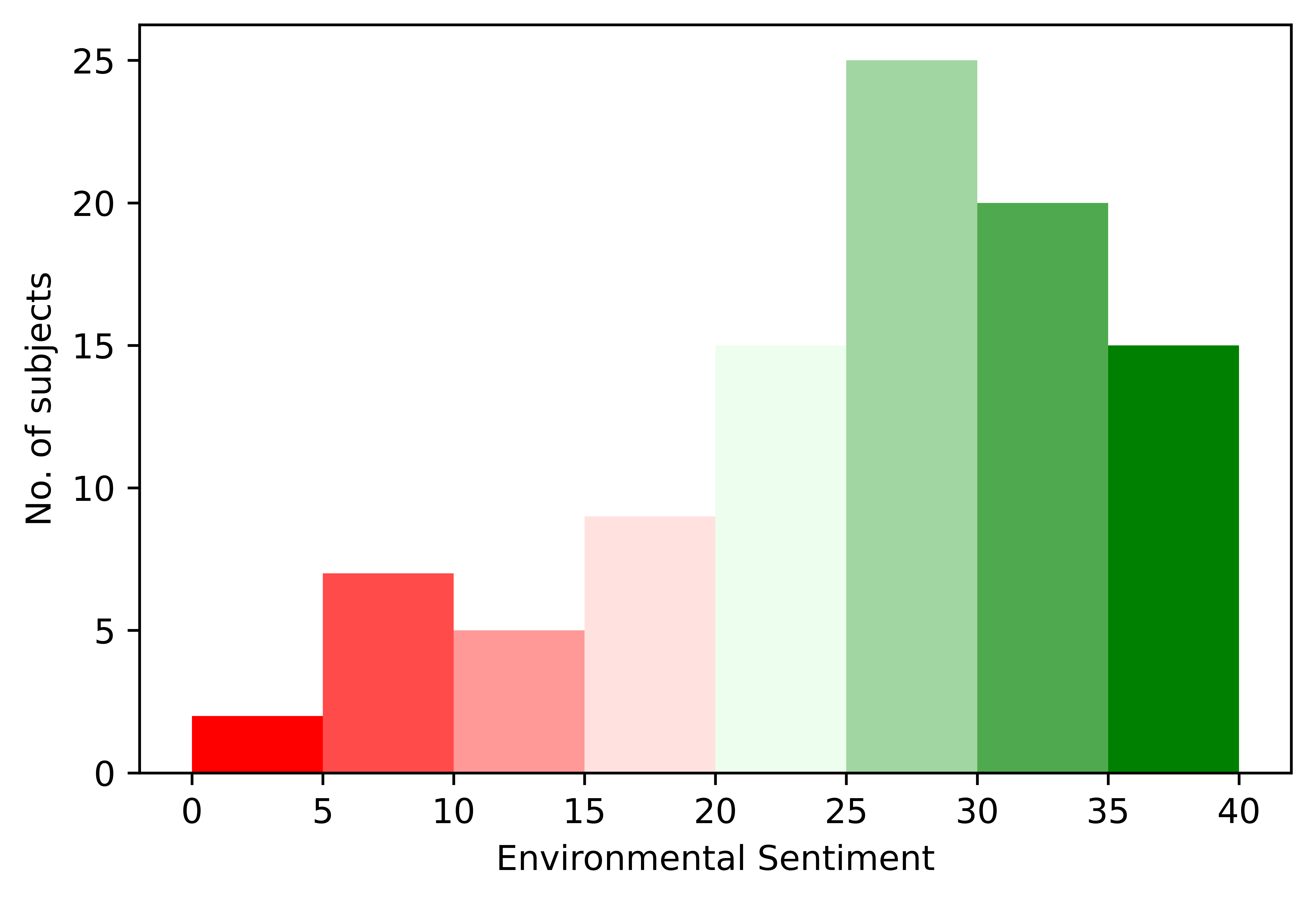} }}%
    \qquad
    \subfloat[\centering]{{\label{fig:attitude-behavior-gap}\includegraphics[width=0.45\columnwidth]{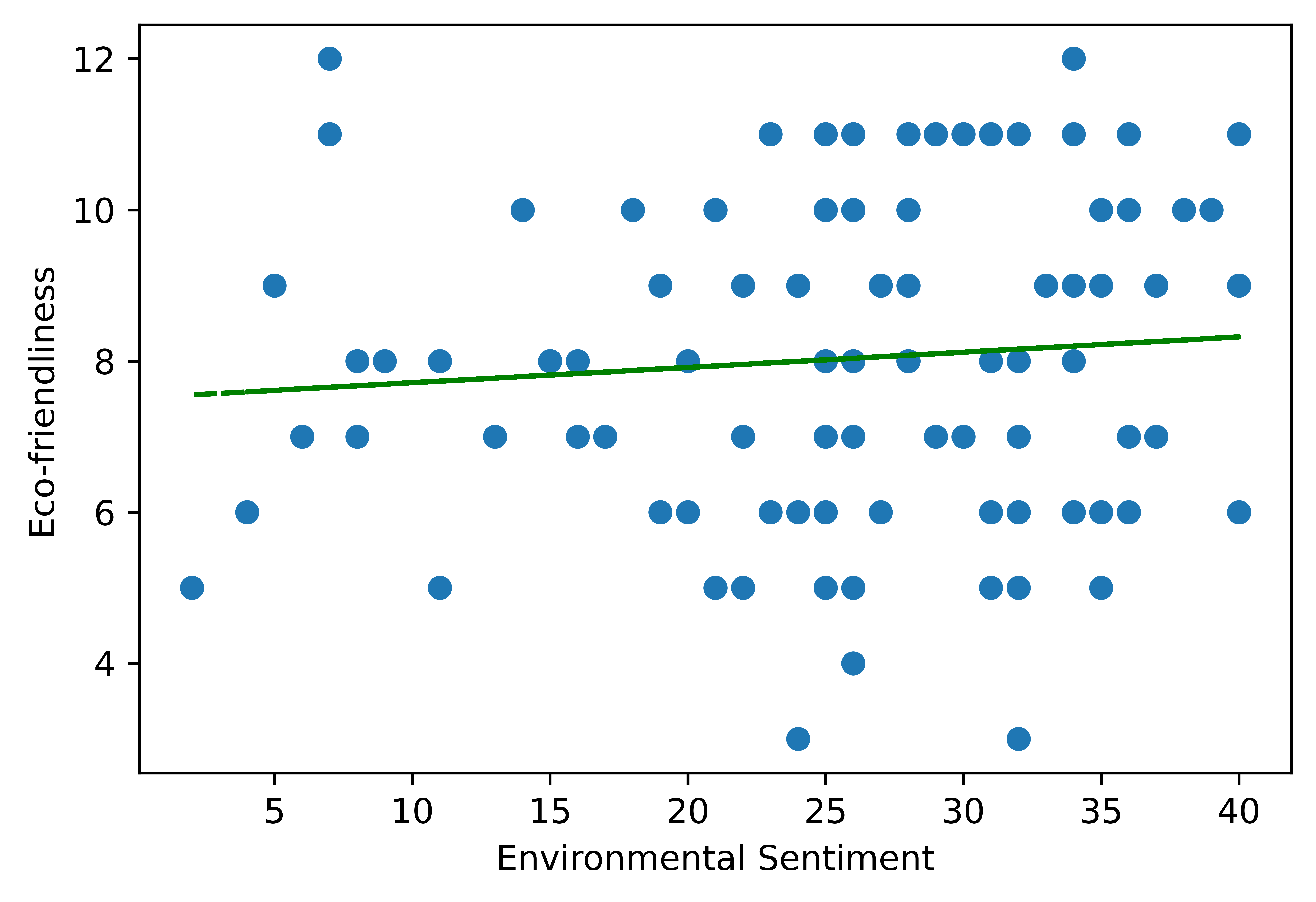} }}%
    \caption{Although people have good intentions, the intentions do not translate well to actual action. (a) shows the environmental sentiment of the subjects (higher value indicates greener sentiment), (b) shows the correlation between environmental sentiment and eco-friendliness as defined by their selection of eco-friendly products ($r = 0.088$; the green line indicates the trend).}%
    \label{fig:attitude-behavior-gap}
\end{figure}

We ask 10 standard questions (Table \ref{tab:sentiment}) to infer the environmental sentiment of all the participants. For each question, we assign a score from 0 to 4 based on the response (strongly disagree: 0, disagree: 1, neutral: 2, agree: 3, strongly agree: 4). So, each participant can score between 0 and 40 -- 20 if they are neutral on average,  above 20 if they demonstrate an eco-friendly attitude, below 20 otherwise. In general, we find that people have good intention (Figure \ref{fig:climate_senti}). 73 subjects (74.5\%) demonstrate an eco-friendly attitude, reporting they are concerned about the environment and intend to take actions to prevent harm while 2 subjects remain neutral, and 23 do not demonstrate such an intention.

\begin{table}[]
\centering
\setlength\tabcolsep{3.2pt}
\renewcommand{\arraystretch}{1}
\resizebox{0.9\columnwidth}{!}{%
\begin{tabular}{llllll}
\hline
\textbf{Environmental sentiment question}                                                                                                                                         & \textbf{SD} & \textbf{D} & \textbf{N} & \textbf{A} & \textbf{SA} \\ \hline
\begin{tabular}[c]{@{}l@{}}I make a special effort to buy paper and plastic products that \\ are made from recycled materials (Q1)\end{tabular}                                   & 7           & 17         & 14         & 41         & 19          \\
I have switched products for ecological reasons (Q2)                                                                                                                              & 8           & 9          & 21         & 41         & 19          \\
\begin{tabular}[c]{@{}l@{}}When I have a choice between two equal products, I purchase \\ the one less harmful to other people and the environment (Q3)\end{tabular}              & 2           & 0          & 12         & 51         & 33          \\
\begin{tabular}[c]{@{}l@{}}I have voted for a candidate in an election at least in part because \\ he or she was in favor of strong environmental protection (Q4)\end{tabular}    & 6           & 11         & 21         & 29         & 31          \\
\begin{tabular}[c]{@{}l@{}}I have avoided buying a product because it had potentially \\ harmful environmental effects (Q5)\end{tabular}                                          & 4           & 7          & 17         & 35         & 35          \\
\begin{tabular}[c]{@{}l@{}}I have read newsletters, magazines or other publications \\ written by environmental groups (Q6)\end{tabular}                                          & 9           & 5          & 14         & 43         & 27          \\
I have signed a petition in support of protecting the environment (Q7)                                                                                                            & 17          & 13         & 17         & 20         & 31          \\
I have given money to an environmental group (Q8)                                                                                                                                 & 21          & 14         & 13         & 27         & 23          \\
\begin{tabular}[c]{@{}l@{}}I have written a letter or called the member of Congress or another\\ government official to support strong environmental protection (Q9)\end{tabular} & 34          & 9          & 13         & 24         & 18          \\
\begin{tabular}[c]{@{}l@{}}I have boycotted or avoided buying the products of a company \\ because I felt that company was harming the environment (Q10)\end{tabular}             & 18          & 11         & 12         & 37         & 20          \\ \hline
\end{tabular}%
}
\caption{Summarized responses of all the participants on the environmental sentiment questions. Here SD, D, N, A, and SA denote ``Strongly Disagree'', ``Disagree'', ``Neutral'', ``Agree'', and ``Strongly Agree'' respectively.}
\label{tab:sentiment}
\end{table}



\subsection{Intention does not always imply action}
We find that consumers' eco-friendly intention rarely translates into action. The eco-friendly behavior of the subjects is measured by the number of eco-friendly products they choose (0-12). We find no statistically significant correlation of gender, age, level of education, or yearly income with subjects' eco-friendliness behavior. This suggests these are either unrelated or may have weak correlation, which would require a larger population to find out. The \emph{attitude-behavior gap} demonstrated in previous literature is also found in our study. As we see in Figure \ref{fig:attitude-behavior-gap}, the correlation between one's environmental attitude (sentiment) and actual eco-friendly behavior is very weak ($r = 0.088$). In addition, we split the participants such that those scoring less than 20 are in the ``less green attitude'' group, and those scoring at least 20 are in ``greener attitude'' group. The behavior is marginally different between these two groups. Using the $t$-test, we fail to reject the null hypothesis that ``there is no relationship between one's greener environmental attitude and actual action'' ($p = 0.75$). Consequently, we fail to establish that greater concern about the environment actually translates to more eco-friendly action ($p = 0.38$).

\subsection{SEER can reduce the attitude-action gap}

In the pre-survey questionnaire, we ask all the participants 5 questions to infer their knowledge about climate change and environmental issues (Table \ref{tab:knowledge}). Participants scoring below the median total score are regarded as part of the ``less knowledgeable'' group and those scoring at least the median are in the ``more knowledgeable'' group. Based on the one-tailed $t$-test, we find that subjects with higher environmental knowledge show more eco-friendly behavior than subjects who are less knowledgeable ($p < 0.0005$). This shows that environmental knowledge can be a powerful tool in increasing purchase of eco-friendly products, which SEER tries to convey to its users.

\begin{table}[]
\centering
\setlength\tabcolsep{3.2pt}
\renewcommand{\arraystretch}{1}
\resizebox{0.9\columnwidth}{!}{%
\begin{tabular}{llllll}
\hline
\textbf{Environmental knowledge question}                                                                                                            & \textbf{SD} & \textbf{D} & \textbf{N} & \textbf{A} & \textbf{SA} \\ \hline
\begin{tabular}[c]{@{}l@{}}I am aware of the environmental and human\\ effects of climate change (Q1)\end{tabular}                                   & 0           & 4          & 13         & 31         & 50          \\ \hline
\begin{tabular}[c]{@{}l@{}}I know that my consumption choices can \\ make a difference (Q2)\end{tabular}                                             & 3           & 5          & 14         & 45         & 31          \\ \hline
\begin{tabular}[c]{@{}l@{}}When I read the description of a product, I can understand\\  whether it is harmful for the environment (Q3)\end{tabular} & 2           & 4          & 27         & 37         & 28          \\ \hline
It is easy for me to recognize an eco-friendly product (Q4)                                                                                          & 1           & 9          & 20         & 43         & 25          \\ \hline
\begin{tabular}[c]{@{}l@{}}I like to read about sustainability, climate change, \\ and the environment often (Q5)\end{tabular}                       & 7           & 10         & 21         & 39         & 21          \\ \hline
\end{tabular}%
}
\caption{Summarized responses of all the participants on the questions asked to infer their knowledge about climate change and environmental issues. Here SD, D, N, A, and SA denote ``Strongly Disagree'', ``Disagree'', ``Neutral'', ``Agree'', and ``Strongly Agree'' respectively.}
\label{tab:knowledge}
\end{table}


\begin{figure}[ht]
\centering
\includegraphics[width=0.60\columnwidth]{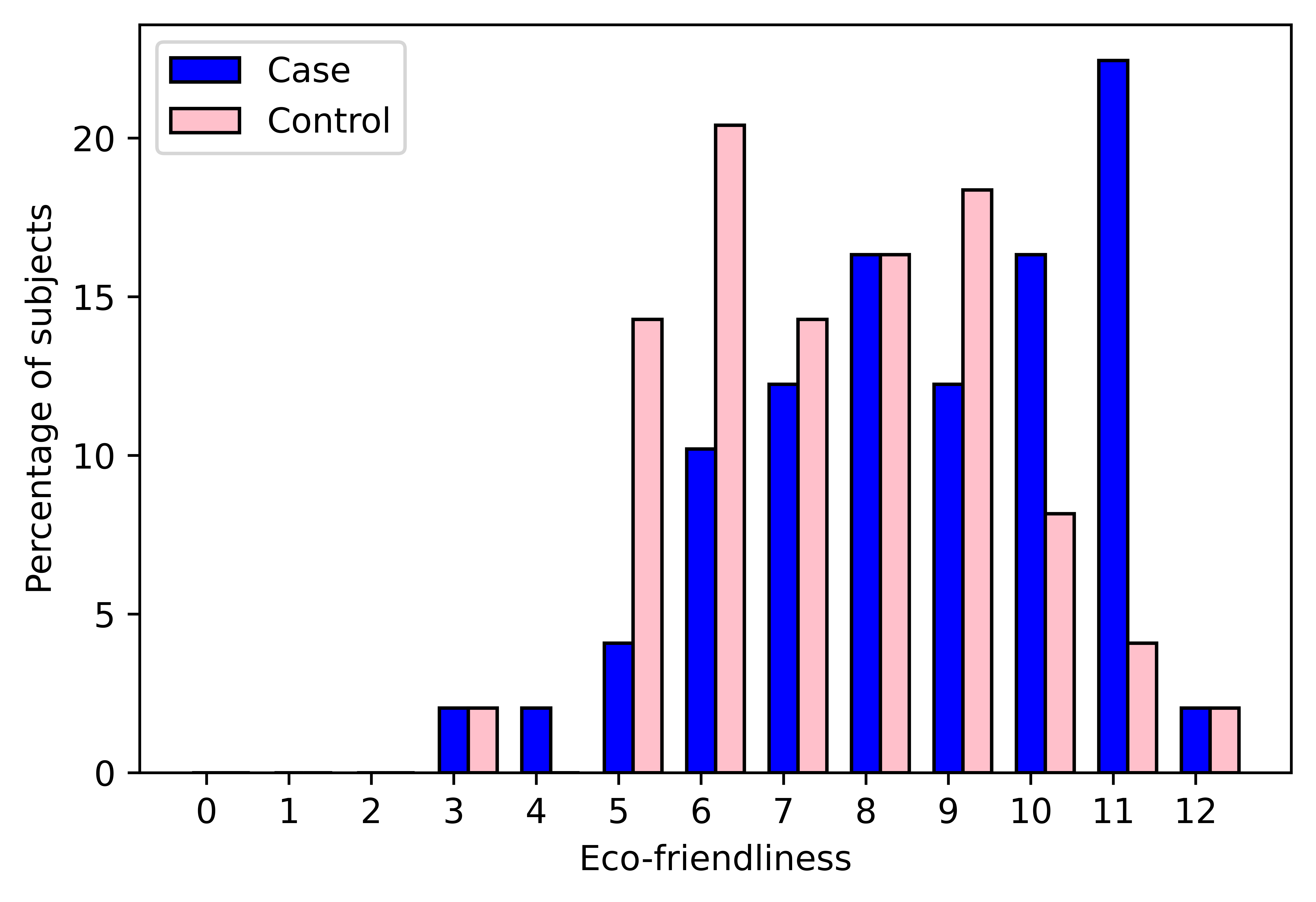}
\caption{The distribution of eco-friendliness in the case and control group. Eco-friendliness is measured by the number of eco-friendly products selected by a participant (out of 12; a higher score indicates more eco-friendly behavior).}%
\label{fig:eco_friendliness}%
\end{figure}

The one-tailed $t$-test reveals that the participants from the case group are significantly more eco-friendly than the control group ($p < 0.005$). Figure \ref{fig:eco_friendliness} shows the eco-friendliness scores of case and control group participants. On average, a subject from the case group selected $8.57$ eco-friendly products (mode = $11$, median = $9$, SD\footnote{SD: Standard Deviation} = $2.13$) while the average for the control group is $7.47$ (mode = $6$, median = $7$, SD = $1.92$). Our experiment suggests that SEER is capable of helping consumers choose more eco-friendly products while shopping online. In the post-study survey, participants from the case group are also asked about their level of agreement on the impact of the SEER prototype and its components on their purchasing decisions. 39 participants (79.6\%) at least agree (agree/strongly agree) that comparing eco-friendliness of products is easy using SEER (29 strongly agreed), 9 participants are neutral, while only one disagrees (no strong disagreement). The proposed prototype not only increases convenience but also increases trust. 30 participants (61.2\%) express that they trust the environmental labels presented in the SEER prototype  more than the traditional websites they use (e.g., Amazon). The above findings indicate that SEER can reduce the ``attitude-behavior'' gap for online shopping by addressing two fundamental barriers: inconvenience and lack of trust in the ``eco-friendly'' labels. 

We also analyze the individual impacts of the proposed three components. 41 participants (83.7\%) express their agreement (28 strongly agree) that the environmental rating makes it easy to identify eco-friendly products, establishing this component as the strongest factor for convenience. The environmental impact summary is primarily responsible for building consumer trust in the provided labels, as 41 participants (83.7\%) at least agree (24 strongly agree) that this component helped them trust the labels. The effect of the environmental keyword highlights is primarily to increase consumer knowledge as it describes what the words/phrases mean and how it is related to the environment. 36 participants (73.5\%) from the case group self-report that they are more aware of the environment after participating in this study, 22 (44.9\%) expressing that they will read more articles related to the environment. 

\begin{figure}%
    \centering
    \subfloat[]{{\includegraphics[width=0.45\columnwidth]{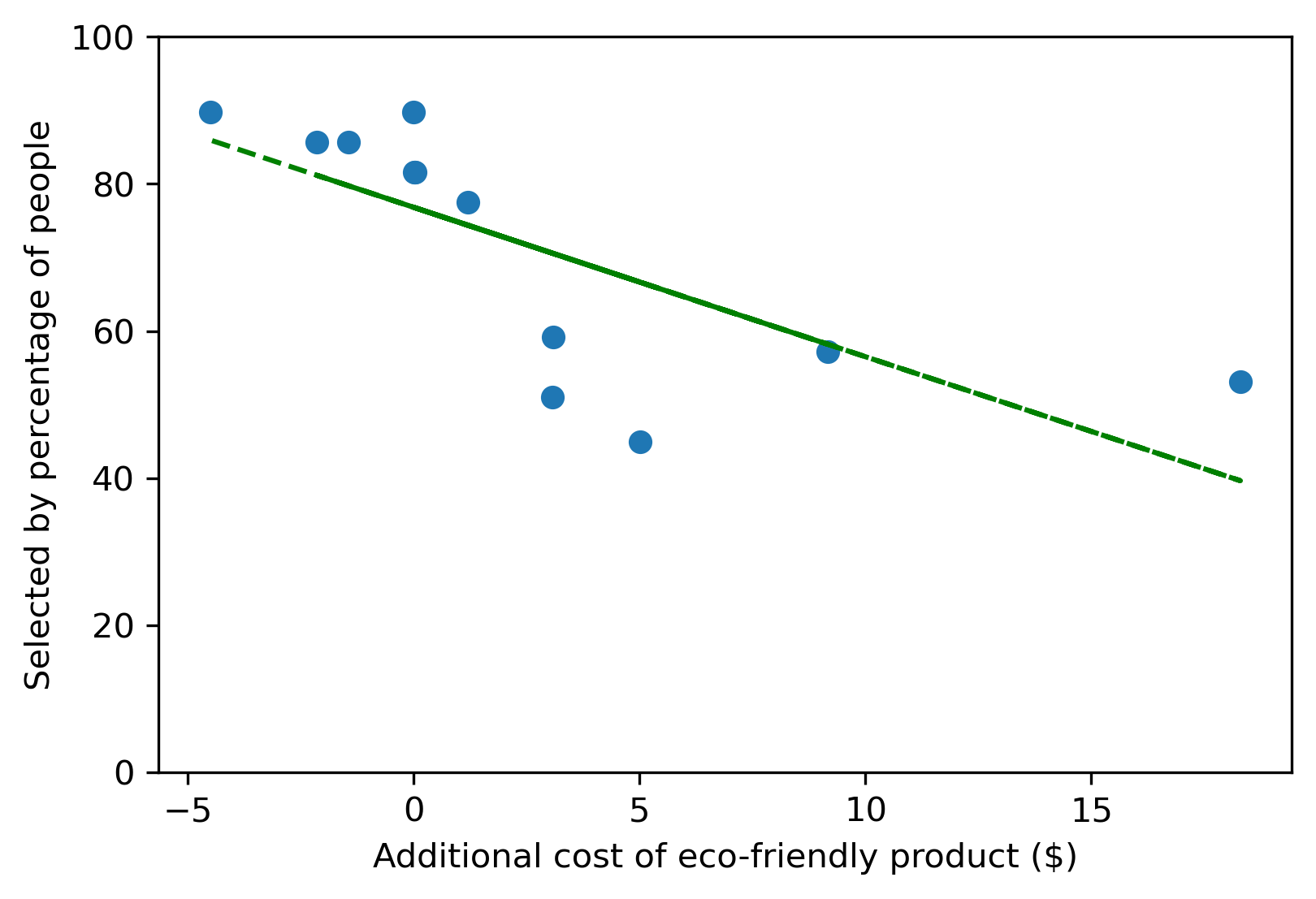} }}%
    \qquad
    \subfloat[]{{\includegraphics[width=0.45\columnwidth]{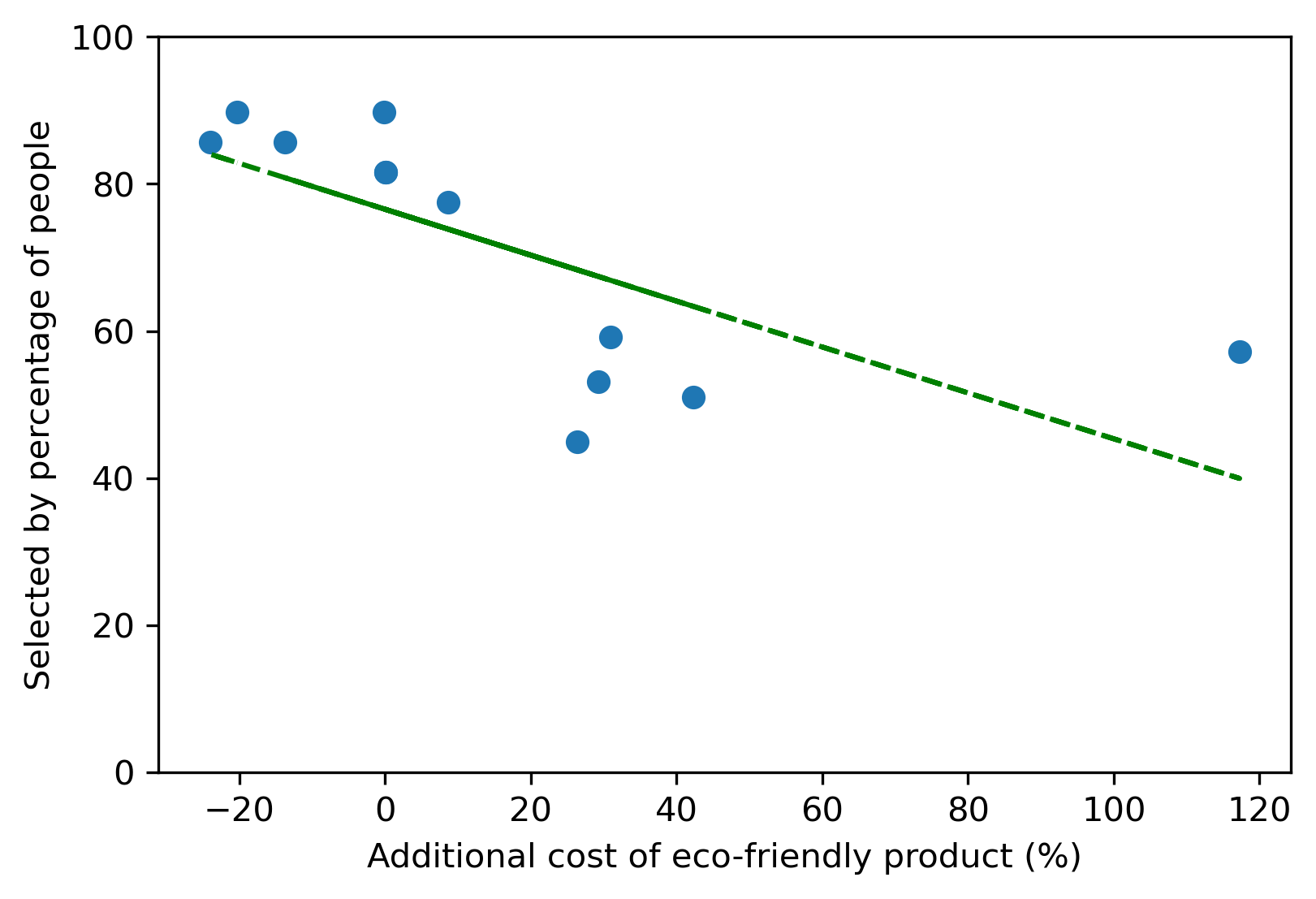} }}%
    \caption{Relation between eco-friendly behavior of the case group and the extra price of the eco-friendly product. The extra price is shown as an exact amount (USD) in (a) and as a percentage in (b). Negative additional cost indicates the eco-friendly product is cheaper, where most of the participants purchase it. The more the additional cost for the eco-friendly product, the less the number of participants purchasing it ($p = 0.007$). The green line indicates the trend.}
    \label{fig:price}%
\end{figure}

\subsection{SEER is user friendly}
To evaluate the usability of the SEER interface, participants from the case group are asked 10 standard System Usability Scale \citep{brooke1996sus} questions (Table \ref{tab:sus}). The average SUS score for SEER is $79.18$ (median = $85$), $68.2$ being the average score for all websites \citep{bangor2009determining}. The mode of the SUS score was surprising -- a perfect score (100), provided by 12 participants (24.5\%). Typically, a SUS score between 70-80 is considered good, more than 80 indicates excellent and less than 50 is not acceptable \citep{bangor2008empirical}. Therefore, we can conclude that SEER has near excellent usability and is acceptable as a user interface.

\begin{table}[]
\centering
\setlength\tabcolsep{3.2pt}
\renewcommand{\arraystretch}{1}
\resizebox{0.9\columnwidth}{!}{%
\begin{tabular}{llllll}
\hline
\textbf{System usability scale questions}                                                                                                 & \textbf{SD} & \textbf{D} & \textbf{N} & \textbf{A} & \textbf{SA} \\ \hline
I think that I would like to use this system frequently (Q1)                                                                              & 2           & 0          & 13         & 39         & 44          \\ \hline
I found the system unnecessarily complex (Q2)                                                                                             & 51          & 14         & 9          & 11         & 13          \\ \hline
I thought the system was easy to use (Q3)                                                                                                 & 0           & 3          & 9          & 23         & 63          \\ \hline
\begin{tabular}[c]{@{}l@{}}I think that I would need the support of a technical person \\ to be able to use this system (Q4)\end{tabular} & 51          & 9          & 14         & 14         & 10          \\ \hline
\begin{tabular}[c]{@{}l@{}}I found the various functions in this system were \\well integrated (Q5)\end{tabular}                         & 0           & 1          & 7          & 46         & 44          \\ \hline
\begin{tabular}[c]{@{}l@{}}I thought there was too much inconsistency in this\\ system (Q6)\end{tabular}                                 & 49          & 12         & 15         & 19         & 3           \\ \hline
\begin{tabular}[c]{@{}l@{}}I would imagine that most people would learn to \\ use this system very quickly (Q7)\end{tabular}              & 0           & 0          & 9          & 30         & 59          \\ \hline
I found the system very cumbersome to use (Q8)                                                                                            & 47          & 10         & 9          & 22         & 10          \\ \hline
I felt very confident using the system (Q9)                                                                                               & 0           & 2          & 12         & 27         & 57          \\ \hline
\begin{tabular}[c]{@{}l@{}}I needed to learn a lot of things before I could \\ get going with this system (Q10)\end{tabular}              & 48          & 8          & 15         & 18         & 9           \\ \hline
\end{tabular}%
}
\caption{Response of all the participants on the system usability scale questions. Here SD, D, N, A, and SA denote Strongly Disagree, Disagree, Neutral, Agree, and Strongly Agree respectively. Odd numbered questions (Q1, Q3, $\cdots$, Q9) are in positive sentiment (SA is the most desired), while the even numbered questions are in negative sentiment (SD is the most desired).}
\label{tab:sus}
\end{table}

\subsection{Eventually, money matters}
However, the price of a product significantly affects consumption behavior \citep{ehrenberg1994after, lichtenstein1993price, ramya2016factors}. As in Figure \ref{fig:price}, we see a significant negative correlation between the extra price a consumer has to pay for an eco-friendly product and the number of consumers who are still willing to pick the eco-friendly product (Pearson's correlation co-efficient $r = -0.73123$ and $p = 0.007$) even while they are using the SEER prototype. On average, more than 80\% of the participants purchase the eco-friendly product when it is cheaper than the non-eco-friendly product or when the prices are similar. This drops to below 60\% when the eco-friendly product is slightly or substantially more expensive. 



\section{Discussion} \label{sec:discussion}
\subsection{Effectiveness of SEER}
Our study establishes that the attitude-behavior gap also holds in an online setting, and that it is indeed possible to nudge people towards eco-friendly purchasing behavior by providing more information regarding the impact of the products. The proposed system SEER aims to assist people who want to make eco-friendly choices but do not end up acting on that sentiment. In our design, we aim to mitigate three specific barriers - lack of trust, lack of knowledge, and inconvenience. From our results, it can be said that addressing these factors indeed increases purchasing of eco-friendly products. In addition, according to the SUS score and thematic analysis of the open-ended feedback from the participants, SEER seems to be well received by the participants. A quote from a participant summarizes our contribution: 

\begin{displayquote}{``I genuinely liked this study a great deal as I thought it was very well-designed, streamlined, easy to interact with, and quite intuitive. Moreover, I liked the conceptual framework of the website showcased in this study as it provided a realistic prototype of a highly usable and user-friendly way of comparing specific parameters of products, which is currently quite difficult, cumbersome, and time-consuming.''}\end{displayquote}

Our thematic analysis of open-ended feedback from participants suggests that people tend to emphasize the price of the product. ``I like the intention to buy at the lowest price, and to save money'' - is what one of our participants said about her purchasing behavior, and it was a sentiment shared by many other participants. Even our quantitative analysis shows similar results - people are less likely to buy the more expensive product when price differences are high, even if it is eco-friendly. However, aligning with prior research, a majority of the participants were willing to pay a little bit extra to purchase the eco-friendly products \citep{gregory2017green, mostafa2016egyptian, wei2018willingness}. 



\subsection{Impact of particular products}

\begin{figure}[ht]
\centering
\includegraphics[width=0.60\columnwidth]{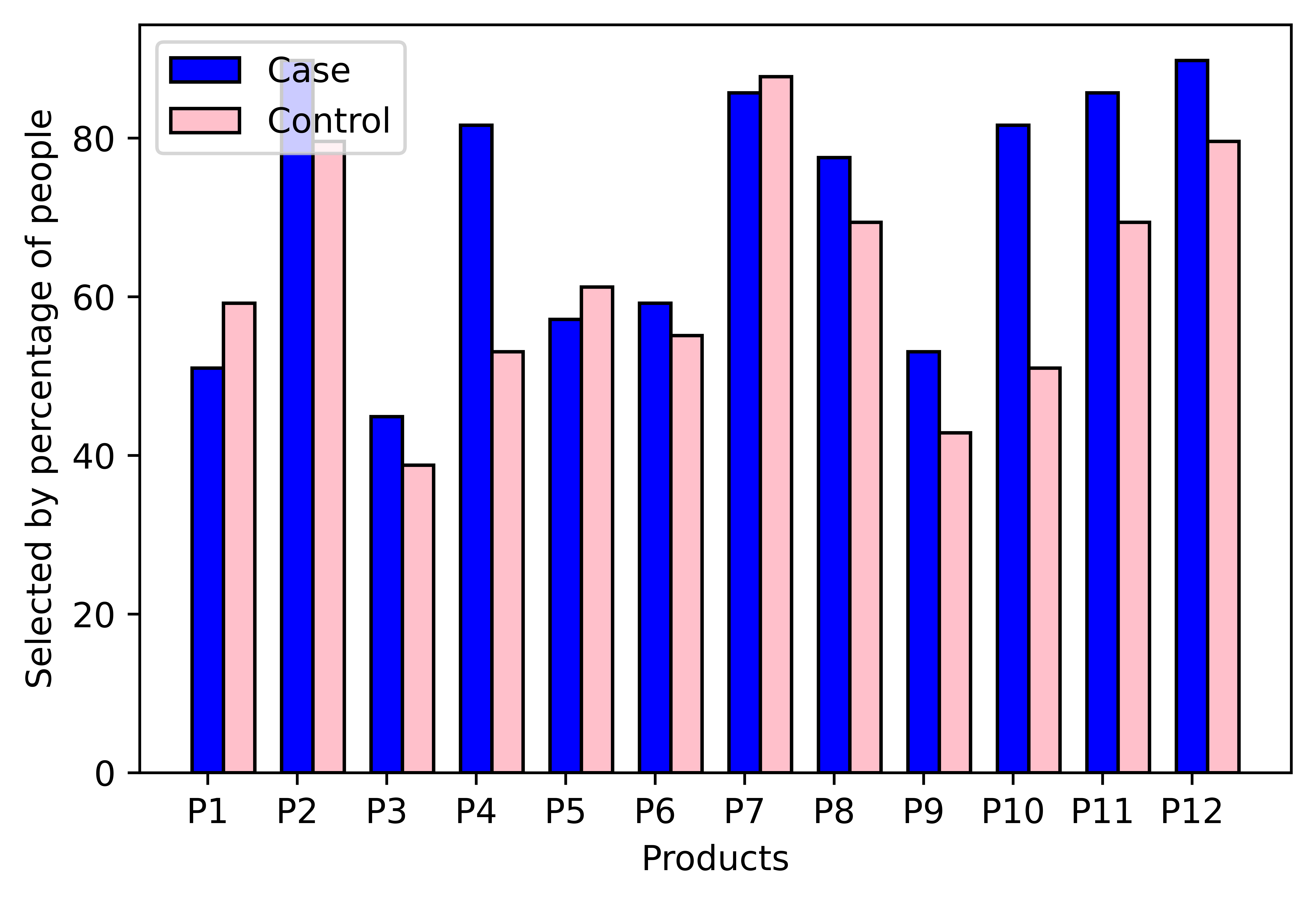}
\caption{Eco-friendliness in the case and control group for all the product types chosen for this study.}%
\label{fig:product_impact}%
\end{figure}

It is possible that people may be more interested in buying eco-friendly for certain types of products than others as shown in Figure \ref{fig:product_impact}. Some eco-friendly products are easier to identify than others, and it is seen that in such cases, both case and control groups almost equally choose the eco-friendly product. For example, in the case of the second selection (P2), one product is a plant-based cleaner while the alternative contains harsh chemicals. Being plant-based is an obvious clue, and most of the subjects (>80\%) from both case and control could identify and purchase the eco-friendly product. Not all the easily identifiable eco-friendly products are popular though -- for P1 (ground decaf, selected by <60\% subjects), one choice is certified organic (thus easily identifiable) but 42\% more expensive than the alternative.  In this case, other factors such as price makes the eco-friendly option less desirable. In general, when identifying the eco-friendly product is difficult, more participants from the case group made the correct choice compared to the control. For example, in the case of P10, one product is a wooden table and another a plastic table. 81.6\% of the subjects from the case group chose the wooden table (more eco-friendly), compared to 51\% in the control group. As reported in the post-study survey, many participants from the control group found it difficult to decide whether trees cut to make the wooden table had more negative environmental impact than the plastic alternative. It is noteworthy that, for both P2 and P10, the price of the eco-friendly and non-eco-friendly products is almost the same, eliminating price as a defining factor for driving consumer decisions.

While we acknowledge that there may be other factors involved, the study is still valuable as a proof-of-concept. Future studies could include more than 12 product types and a deeper analysis of the relationship between product attributes and the receptivity of eco-friendly products. 

\subsection{Policy implications}

In addition to directly informing about specific purchase decisions, such a system could be used to provide additional incentives or rewards for green purchasing. For example, people could be given an “eco-friendliness score” or earn “green points” depending on their purchasing history, and get rewarded based on it. Providing eco-friendly ratings may also encourage corporations to be more transparent about the materials they use in their products. If sales of eco-friendly products indeed increase with increased awareness, a positive feedback loop may be created, driving companies to follow more eco-friendly practices. 

However, it must be noted that the effectiveness, efficiency, and equity of such a system depends on implementation details.  For example, if companies self-rate their products, the system could exacerbate “greenwashing" \citep{laufer2003social}. Companies may provide false information, omit harmful ingredients or practices, or use more eco-friendly keywords to inflate their products’ eco-friendliness scores.  It is also important to recognize that because certain eco-friendly products tend to be more expensive, any rewards associated with green purchasing patterns could exacerbate economic disparities, as lower-income consumers might not be able to take advantage of these rewards. However, companies could tie this system to justice-promoting strategies, such as making donations to climate justice or community environmental organizations with a percentage of profits from eco-friendly purchases. It must be noted that SEER does not directly address or mitigate underlying social/environmental/economic issues related to climate change; it is simply a tool to provide consumers with more information to aid in making more eco-friendly choices. Therefore, the actual impact of our system depends on actions at the policy level such as eco-friendly rating schemes, incentive structures, and tie-in policies by participating companies.

\subsection{Integrating SEER components as optional information}
Some participants believe that individuals cannot prevent climate change, and using a system like SEER would place the responsibility of climate change on individuals. Moreover, it can cause emotional stress for individuals who cannot afford to be eco-friendly as they can see the negative impact of their choices on the environment. To address these issues, the components proposed in SEER can be made optional so it does not burden the consumers with information they do not want.

\subsection{Immediate and long term impact}
The study had an immediate impact on the subjects by increasing environmental awareness among participants. Many reported learning more about the environment, and expressed more willingness towards eco-friendly consumption in our post-study survey:

\begin{displayquote}
``Thanks for the reminder, sometimes I feel like individuals cannot do much to help. I think it is easier to blame big companies for not doing enough, but it is true that consumer pressure can make a difference.''

``It is a humble reminder to think about the environment whenever we want to make a purchase''

``I learned more about eco-friendly products and the effects of climate change. I will try to focus on buying more green products from now on.''
\end{displayquote}

While it is challenging to generate reliable environmental ratings and explanations, our proposed idea can potentially have a huge impact in tackling climate change by reducing carbon emission from potentially every online purchase. According to the U.S. Department of Commerce, e-commerce retail sales for the country accounted for more than 13\% of total retail sales in 2021 (First - Third quarter) \citep{estquarterly2021}. Earlier research shows that with appropriate intervention like reminding consumers of eco-friendly behaviors, consumers' carbon footprint from food and drink purchases alone can be reduced by 3 kg a week on average \citep{panzone2018impact}. In the United States, almost 25\% people (about 80 million) shop online at least once in a month \citep{optinmonster2021}. If one online purchase has a carbon footprint similar to a single grocery trip (a modest assumption), it is possible to reduce 2.88 megatons of carbon emission annually if every U.S. online shopper uses SEER ($80 \text{ million monthly shoppers} \times 12 \text{ months} \times 3 \text{ Kg carbon reduction per monthly shopping}$). This is equivalent to more than half a million people getting rid of their car, considering an average passenger vehicle emits 4.6 tons of $CO_2$ per year \citep{epa2021vehicle}.

\section{Limitations and Future Work} \label{sec:future_works}
A major limitation of our study is that the environmental impacts of the products are simply hand-annotated by an expert. However, in the real world, there are thousands of products, and it would be very difficult to manually annotate them all. Moreover, as the product life-cycle is constantly changing, product ratings must be updated accordingly. This naturally brings into question -- how do we generate environmental-impact information in the real world? One method could be delegating this task to the manufacturers and an independent, neutral group could monitor it. This approach brings consumer trust into question, as consumers may not trust environmental labels provided by the manufacturers \citep{delmas2011drivers}. Another method can be an autonomous body manually generating these ratings for the most popular types of products first, and then increase coverage with time. But, this might be difficult to scale and continuously update in the light of new knowledge. One promising method seems to be crowd-sourcing, inspired by the success of Wikipedia -- the encyclopedia that maintains nearly 4 million articles using crowd-sourcing, and remains as accurate as Encyclopedia Britannica \citep{giles2005special}. Finally, with the advances in machine learning (e.g., natural language processing), it might be possible to generate the environmental ratings and impact statements automatically. For example, a knowledge graph that captures environmental entities and their relations \citep{hogan2021knowledge, mishra2021neuralnere} might help automatically identify the related keywords and generate explanations to how those keywords interact with the environment.

In our study, the motivational climate video acted as an emotional incentive for green consumption, similar to emotional green advertising used in prior research \citep{matthes2014consumers}. Priming both case and control groups in this way allows us to probe into the question - if someone wants to make eco-friendly purchases, how effective would our proposed design be compared to the traditional ones? In this context, priming the participants is appropriate. Future studies could be done without priming the participants to simulate a more real-life scenario, where not all people are always actively thinking about the environment.

Another potential limitation of the study is the sample size and the demography of the population. Scaling up the experiment would be challenging as it would significantly increase the cost incurred in the experiment. Future studies could be done in an entertaining way (e.g., gamification) so users will be willing to participate for free. Alternatively, we could partner up with an e-commerce business and present our interface to a larger number of users. Moreover, all the subjects are from the United States, where literacy rate and per capita income are higher compared to many other countries. Future studies are needed to assess how receptive people will be to SEER in other countries where literacy rate is lower and/or there are financial constraints. 

Our study is a proof-of-concept that shows SEER is an effective, easy-to-use, and well-received prototype for people willing to be eco-friendly. We propose a new interdisciplinary research direction that connects user interface design with sustainable consumption research. However, we also acknowledge that this is a complex issue with other major contributing factors such as price, availability, personal and social norms, habit, and so on, which are not addressed by the study. Despite the limitations, we hope our study will inspire future research for e-commerce re-design, and encourage interdisciplinary research for promoting sustainability.

\section*{Ethics}
All the participants who completed the study received a payment of 10 USD, and an additional bonus by saving their budget. The participants, at any point, were permitted to quit the study without facing any consequence. After the end of the study, all the participants were clarified that the products they purchased are not for any real school, and their choices did not cause any harm to the environment, considering some participants might feel guilty of purchasing non-eco-friendly products. The entire study was approved and supervised by the Institutional Review Board (IRB) of the University of Rochester. All researchers who conducted the study and analyzed data are certified by IRB to conduct human behavior studies.

\section*{Acknowledgement} \label{sec:ack}
We thank Goergen Institute for Data Science for the Seed Funding of this project. We also acknowledge Liudmila Paymukhina for her valuable feedback on formulating the experiment.

\section*{Code and Data Availability}
Upon acceptance, both code and data will be made publicly available following the IRB protocol of this project.

\section*{Author contributions statement}
M.S.I, A.M.P., C.W., K.B., and E.H. designed the experiment and the prototype; C.W., V.K., and S.U. implemented the prototype; M.S.I., A.M.P., and C.W. conducted the experiments; M.S.I. analyzed the results; M.S.I., A.M.P., and E.H. wrote the manuscript; K.B. and K.K. helped improve the manuscript; E.H. supervised the project. All authors reviewed the manuscript.

\nolinenumbers

\bibliographystyle{elsarticle-harv}
\bibliography{main}

\end{document}